\documentclass[prl, aps, twocolumn, superscriptaddress, showpacs]{revtex4}

\usepackage{slashed}
\usepackage{graphicx}
\usepackage{subfigure}
\usepackage[usenames, dvipsnames]{color}
\usepackage{graphics}
\usepackage{hyperref}
\usepackage{bm}
\usepackage{amsmath}
\usepackage{color}
\usepackage{amsfonts}
\hypersetup{backref,
colorlinks=true,
linkcolor=blue,
linktoc=page,
citecolor=blue}

\newcommand{\pr}[1]{\ensuremath{\left[#1\right]}} 
\newcommand{\pc}[1]{\ensuremath{\left(#1\right)}}

\newcommand{\md}[1]{\ensuremath{\left\vert#1\right\vert}}

\begin{document}

\title{Quantum Thermal Machines Fuelled by Vacuum Forces}

\author{Hugo Ter\c{c}as}
\affiliation{Physics of Information and Quantum Technologies Group, Instituto de Telecomunica\c{c}\~oes, Lisbon, Portugal}
\email{hugo.tercas@it.lx.pt}
\author{Sofia Ribeiro}
\affiliation{Physics of Information and Quantum Technologies Group, Instituto de Telecomunica\c{c}\~oes, Lisbon, Portugal}
\author{Marco Pezzutto}
\affiliation{Physics of Information and Quantum Technologies Group, Instituto de Telecomunica\c{c}\~oes, Lisbon, Portugal}
\affiliation{Instituto Superior T\'ecnico, Universidade de Lisboa}
\author{Yasser Omar}
\affiliation{Physics of Information and Quantum Technologies Group, Instituto de Telecomunica\c{c}\~oes, Lisbon, Portugal}
\affiliation{Instituto Superior T\'ecnico, Universidade de Lisboa}


\begin{abstract}

We propose a quantum thermal machine composed of two nanomechanical resonators (NMR) (two membranes suspended over a trench in a substrate), placed a few $\mu$m from each other. The quantum thermodynamical cycle is powered by the Casimir interaction between the resonators and the working fluid is the polariton resulting from the mixture of the flexural (out-of-plane) vibrations. With the help of piezoelectric cells, we select and sweep the polariton frequency cyclically. We calculate the performance of the proposed quantum thermal machines and show that high efficiencies are achieved thanks to (i) the strong coupling between the resonators and (ii) the large difference between the membrane stiffnesses. Our findings can be of particular importance for applications in nanomechanical technologies where a sensitive control of temperature is needed.
\end{abstract}
\maketitle
{\it Introduction.}
The relation between the basic principles of quantum mechanics and those of thermodynamics constitute a fundamental question which is not yet completely understood \cite{review1, review2}. When dealing with miniaturized systems, quantum effects come into play: fluctuations are no longer just thermal in their origin but quantum, i.e. they appear even at zero temperature and are ubiquitous in quantum fields. As a consequence, it is not clear why the time-reversible, unitary dynamics that describes quantum processes should lead to a system ever reaching thermal equilibrium. Quantum-mechanical effects are then expected to play a crucial role on the transport properties of heat in such conditions. For example, it is known that the Fourier law for heat transfer is violated for nanomechanical resonators (NMR) \cite{fourier_violation}. Moreover, the possibility of exploiting the features of quantum mechanics (coherence, entanglement) to build quantum heat engines and refrigerators may lead to useful applications in quantum technologies \cite{quantum_tech_app}. A natural challenge is therefore to understand the features of quantum thermal machines (QTM) within the accessible technology. QTMs have been the subject of intense theoretical work within the last decades (see e.g. \cite{scovil, scully, humphrey, scully2, kieu, dillenschneider, gemmer} and Ref. \cite{hanggi} for a review). Moreover, realizations of quantum heat engines (QHEs) have been put forward in experiments with single ions \cite{abah_single_ion}, cold gases \cite{brantut} and optomechanical setups \cite{zhang}. Extracting work from entanglement has also been considered in the context of quantum information \cite{oppenheim, vedral, rossnagel, ciampini}. Quantum refrigerators (QR) have also been investigated at the nanoscale, and the fundamental limits to their performance have been determined \cite{linden2010}. 

On the other hand, in the run for smaller and more compact technology, fluctuations of the electromagnetic (EM) vacuum start to play a crucial role. Vacuum forces, as resulting from the quantization of the electromagnetic spectrum  yielding to the so-called Casimir interaction \cite{casimir, lifshitz}, are particularly important, as they are very sensitive to small distance variations. Those forces are often dominant and can overpower applied forces and can be used to actuate and sense very small mechanical displacements. The control of vacuum forces thus provides a plethora of applications, such as atom trapping \cite{chang2014}, gravity metrology \cite{onofrio}, mechanical sensing \cite{muschik, rodrigues2008}, corrugated-surface microscopy \cite{moreno2010}, and nanosphere levitation \cite{levitationCasimir}. Can we then explore vacuum forces to fuel quantum thermal machines?
\begin{figure}[t!]
\includegraphics[scale=1]{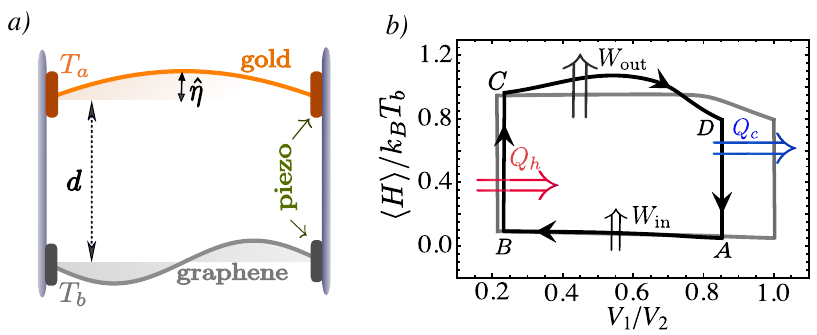}
\caption{(color online) a) Schematic representation of the  quantum thermal machine. Two nanomechanical resonators
(gold and graphene membranes) are clamped in a dilution chamber at temperatures $T_a$ and $T_b$. A piezoelectric cell selects the flexural (out-of-plane) mode and controls the thermodynamical cycle. b) Diagrammatic representation of the quantum heat engine based on the lower polariton mode. Two adiabatic ($\mathcal{Q}^*=1$, lighter curve) or quasi-adiabatic ($\mathcal{Q}^*=1.2$, darker curve) isentropic strokes (compression and expansion), and two isochoric strokes are represented.  $T_a= 1$ mK and $T_b=10$ mK. A quantum refrigerator is obtained by inverting the direction of the cycle.}
\label{fig_cycle}
\end{figure}

In this paper, we propose a QTM based on an interface between two NMRs. The energy associated to the out-of-plane (flexural) phonons can be harnessed to produce work if the two sheets are separated from a few $\mu$m. In that case, EM vacuum fluctuations fuel a thermal machine operating between two thermal reservoirs of flexural phonons. Our setup consists of a gold and a graphene membranes kept at different temperatures in a cryogenic environment. Piezoelectric cells will act as a piston, allowing us to select the flexural modes of and to sweep them cyclically. By reversing the direction of the cycle and changing the membrane temperatures, our system can work either as a QHE or as a QR. We then define the thermodynamical cycles and analyse their performances. For a particular, yet tunnable, temperature ratio, we observe leading efficiencies at maximum power of the order of 70\%. We conclude that the high-efficiency of these QTMs is a hallmark of NMRs thanks to (i) the strong coupling (hybridization) between the flexural modes and (ii) the large difference between the membrane stiffnesses. 
\par{\it Casimir interaction at cryogenic temperatures.}
%
The Casimir energy per unit surface at zero temperature is given by \cite{casimir, lifshitz, supp}%
\begin{equation}
E(d) \simeq \frac{\hbar}{4 \pi^2} \sum_{j=\mathrm{TE,TM}}\int_{0}^\infty k dk \int_0^\infty d \xi \ln \pr{f_k^j \pc{i \xi}},
\label{eq:EnergyQuantum}
\end{equation}
where ${\bm k}$ is the in-plane wavevector, $\xi=i\omega$ and $\omega$ is the frequency of the transverse-electric (TE) and transverse-magnetic (TM) EM modes. For the geometry of Fig.~\ref{fig_cycle}, $f_k^j = 1 - \exp \pc{-2 \kappa_2 k d} r_{21}^j r_{23}^j$, with $r_{mn}$ being the reflection coefficient for a wave on the interface between medium $m$ and $n$, $d$ is the distance between the membranes, and $ \kappa_n = \sqrt{1+\varepsilon_n \pc{i \xi} \pc{\xi/c k}^2}$ \cite{supp}. Here, $\varepsilon_n$ represents the dielectric function of the medium $n$ and $c$ is the speed of light. Finite-temperatures effects are taken into account by replacing the second integral in Eq.~\eqref{eq:EnergyQuantum} by a sum over the Matsubara frequencies \cite{sofiaPRA2013, supp}. At cryogenic temperatures of few mK \cite{song2014}, thermal contributions are very weak \cite{supp}. This allows us to safely neglect out-of-equilibrium corrections resulting from the resonators being kept at different temperatures \cite{Antezza}. A numerical fit to Eq.~\eqref{eq:EnergyQuantum} provides $E(d)\simeq C_3/d^3$, with $C_3=-1.14 \times 10^{-11}$ Jm.

{\par \it Kirchhoff-Love theory of mechanical vibrations.}
%
The dynamics of suspended membranes is described by the Kirchhoff-Love plate theory \cite{amorim, tercas2015}. The Lagrangian density of the $j-$plate ($j=a, b$) can be written as $\mathcal{L}_j=\rho_j \dot {\bm \eta}_j^2/2-\mathcal{H}_j$, where $\rho_j$ is the areal mass density and ${\bm \eta}_j({\bm x})= \eta_j({\bm x}){\bm e}_z$ is a continuous vector field describing the vertical (out-of-plane) displacement of the $j$th membrane at position ${\bm x}=(x,y)$. The energy density can therefore be expressed as
\begin{equation} 
\mathcal{H}_j=\frac{1}{2}D_j\left(\nabla^2{\bm \eta}_j\right)^2+ \gamma_{j}\nabla^2 {\bm \eta}_j ,
\label{Eq_hamilton1}
\end{equation}
where $D_j$ is the bending stiffness, $\gamma_{j}$  is the clamping tension with the substrate (here considered to be isotropic), and $\nabla^2{\bm \eta}_j$ is the local curvature. Quantization of the flexural modes is provided by the prescription $\bm{\eta}_j({\bm x})=\mathcal{S}^{-1}\sum_{\bm k,\sigma} h_{\bm k}^{(j)} e^{i{\bm k}\cdot{\bm x}}(c_{k,\sigma}^{(j)}+c_{k,\sigma}^{(j)\dagger}) {\bm e}_\sigma$, where $\sigma$ is the mode polarization, $\mathcal{S}$ the plate area, $h_{\bm k}^{(j)}=[\hbar /(M_j \omega_{\bm k}^{(j)})]^{1/2}$ is the mode amplitude, $M_j = \mathcal{S} \rho_j$ is the $j$th resonator mass and $c_{{\bm k},\sigma}^{(j)}$ is the usual bosonic operator satisfying the commutation relation $\left[ c_{{\bm k},\sigma}^{(j)}, c_{{\bm k}',\sigma'}^{(j')\dagger}\right]=\delta_{{\bm k},{\bm k}'}\delta_{\sigma,\sigma'}\delta_{j, j'}$. The single-particle Hamiltonian therefore reads $H_0=\hbar \sum_{k,\sigma} \left(\omega_{k}^{(a)} a_{k,\sigma}^\dagger a_{k,\sigma} + \omega_{k}^{(b)} b_{k,\sigma}^\dagger b_{k,\sigma} \right)$, where we made the identification $c_{k,\sigma}^{(a)}=a_{k,\sigma}$ and $c_{k,\sigma}^{b}=b_{k,\sigma}$, with $k=\vert {\bm k} \vert$ for symmetry reasons. Finally, the bare frequencies simply read $\omega_k^{(j)}=\sqrt{D_j k^4+\gamma_j k^2}$.  

{\par \it Proximity field approximation: the effect of the flexural modes.}
%
\begin{figure}[t!]
\includegraphics[width=0.95\columnwidth]{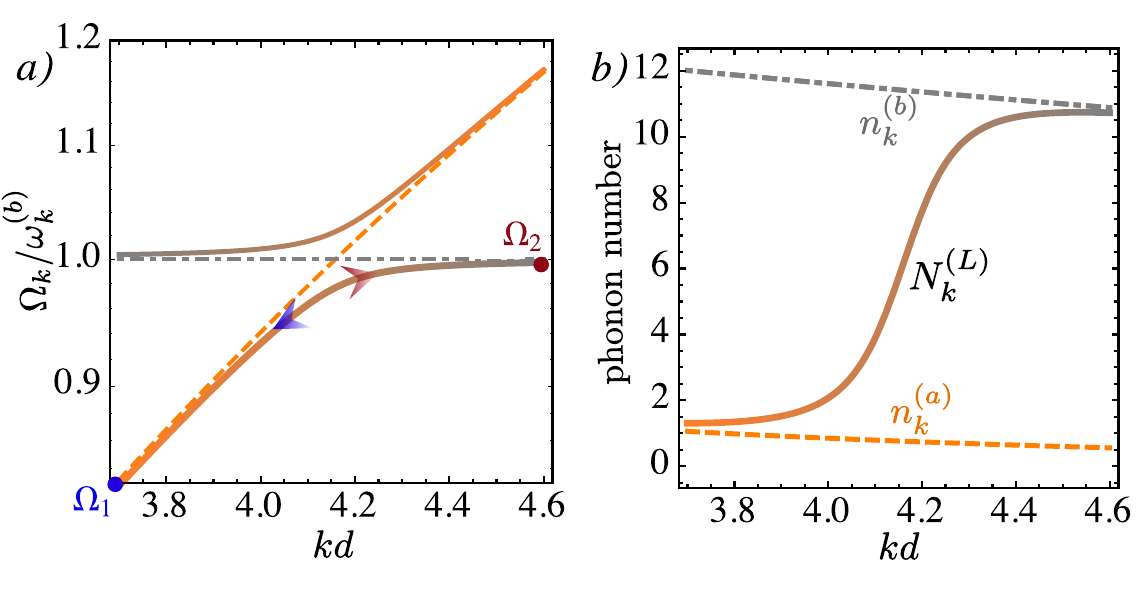}
\caption{(color online) a) Spectrum of Eq. (\ref{Eq_hamilton2}) near the avoided-crossing region. Golden/dashed (resp. gray/dot-dashed) line represents the bare gold, $\omega_k^{(a)}$ (resp. graphene, $\omega_k^{(b)}$) flexural dispersion. Solid lines are the polariton modes and $\Omega_1$ ($\Omega_2$) is the initial (final) mode used in the thermodynamical cycles. b) Bare ($n_k^{(a)}$ and $n_k^{(b)}$) and lower-polariton ($N_k^{(L)}$) phonon number for $T_a=1$ mK and $T_b=10$ mK (see \cite{supp} for details).}
\label{fig_dispersion}
\end{figure}
The out-of-plane motion induces fluctuations in the Casimir potential. Making use of the proximity-field approximation \cite{derjaguin1934, thorsten2001, gies2006, rodriguez2011}, the local effects of a ripple with wavelength $\lambda$ can be taken into account in Eq.~(\ref{Eq_hamilton1}) provided $\mathcal{H}_j\rightarrow \mathcal{H}_j +\sum_i \mathcal{U}_{ij}$, where
\begin{equation}
\mathcal{U}_{ij}(d)\simeq \frac{C_3}{d^3}\left[ 1+k\frac{\eta_i^\dagger \eta_j}{ d}\right]. 
\label{fluctuation}
\end{equation}
With the quantization prescription above, the latter yields the following interaction Hamiltonian $H_{\rm int}=\sum_{k}g_k a_k^\dagger b_k +{\rm h.c.}$, where $g_k=2 C_3 k h_{k}^{(a)}h_{k}^{(b)}/d^4$ is the coupling strength (the factor of 2 appears due to the summation over the polarization index $\sigma$). Here, we made use of the rotating-wave approximation, which amounts to neglect the terms $a_k b_k$ and $a_k^\dagger b_k ^\dagger$ that do not conserve the total number of excitations. Such an approximation is well justified for $g_k\ll \omega_k^{(a,b)}$, a fact that will be verified {\it a posteriori}. The potential fluctuation in Eq.~(\ref{fluctuation}) also induces corrections (AC Stark shifts) to the bare frequencies as $\omega_k^{(j)}\rightarrow \omega_k^{(j)}+2C_3 k \vert h_k^{(j)}\vert ^2/d^4$, such that the total Hamiltonian $H=H_0+H_{\rm int}$ reads
\begin{equation}
H=\hbar \sum_k \left(\omega_k^{(a)} a_k^\dagger a_k +\omega_k^{(b)}b_k^\dagger b_k\right)+\sum_k g_k a_k^\dagger b_k.
\label{eq_hamilton_pol}
\end{equation}
Eq.~\eqref{eq_hamilton_pol} can be diagonalized by introducing the polariton operators $A_k=u_k a_k+v_k b_k$ and $B_k=v_k b_k-u_k a_k$, where the Hopfield coefficients satisfy the normalization condition $\vert u_k\vert^2+\vert v_k\vert^2=1$. In the new basis, the Hamiltonian reads
\begin{equation}
H=\hbar \sum_k\left( \Omega_{k}^{(L)} A_k^\dagger A_k+\Omega_{k}^{(U)} B_k^\dagger B_k\right),
\label{Eq_hamilton2}
\end{equation}
where the lower (L) and upper (U) polariton frequencies are given by $$\Omega_{k}^{(U,L)}=\frac{1}{2}\left( \omega_k^{(a)}+\omega_k^{(b)} \pm \sqrt{\left(\omega_k^{(a)}-\omega_k^{(b)} \right)^2+4\vert g_k\vert ^2 }\right).$$ A measure of the quantum coherence between $a$ and $b$ modes is given by the Rabi frequency $\Lambda\equiv 2 g_{k_c}$, where $k_c$ is the avoided-crossing of frequency $\omega_c\equiv \omega_{k_c}^{(a)}=\omega_{k_c}^{(b)}$. The features of Eq.~(\ref{Eq_hamilton2}) are depicted in Fig.~\ref{fig_dispersion}. Strong-coupling is achieved if $\Lambda$ is much larger than the decoherence rate $\Gamma$. Since the flexural modes are extremely long-lived at cryogenic temperatures \cite{jiang2015}, the polariton decay rate is essentially attributed to polariton-flexural phonon scattering \cite{supp}. For the case of a graphene-gold interface, we obtain $\Lambda \sim 5$ MHz and $\omega_c ~\sim 100$ MHz, and therefore the strong coupling condition $\Lambda \gg \Gamma$ typically holds.  

{\par \it Quantum Heat Engine.}
%
\begin{figure}[t!]
\includegraphics[scale=0.34]{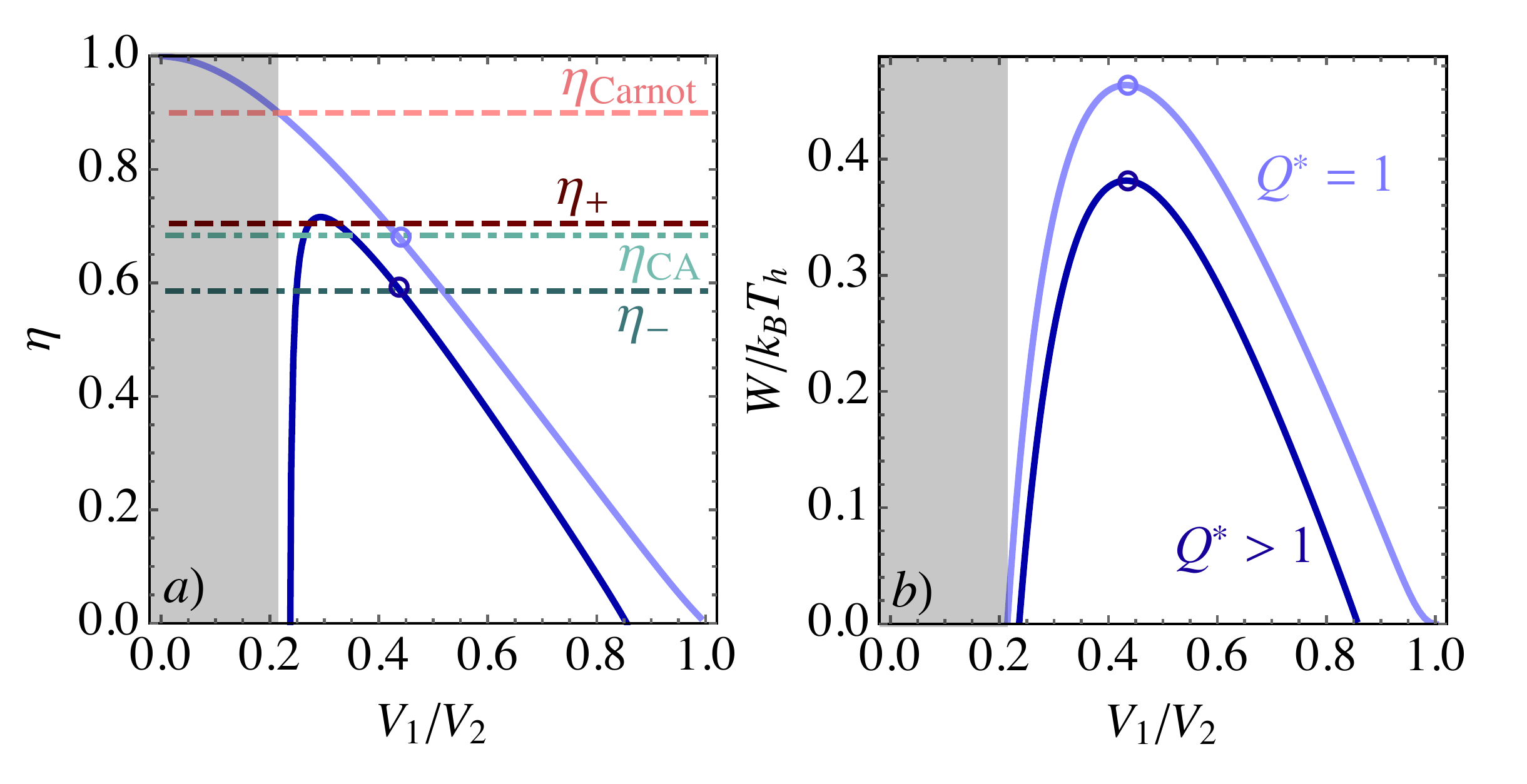}
\caption{(color online) Efficiency and power of the QHE. a) Efficiency for a cycle operating with two adiabatic ($\mathcal{Q}^*=1$, lighter blue) and quasi-adiabatic ($\mathcal{Q}^*=1.005$, darker blue) isentropic strokes (compression and expansion). The performance of the machine is upper-bounded by the Carnot efficiency. b) Work extracted during the thermodynamical cycle. The circles represent the compression ratio for which the power is maximum and the shadowed area represents the region of work reversion ($W<0$). The efficiency at maximum power is lower-bounded by the Curzon-Ahlborn (CA) efficiency for the adiabatic case and by the value $\eta_-$ discussed in the text for quasi-adiabatic case. The cycle amplitude is $\Omega_2/\Omega_1\simeq 1.32< T_2/T_1\simeq T_b/T_a$. $T_a=1.0$ mK and $T_b=10$ mK.}
\label{fig_Eff}
\end{figure}
We now construct a quantum heat engine, based on the Otto cycle, for which the lower polaritonic mode is the working ``fluid". The thermodynamical cycle consists of four strokes, as represented in Fig.~(\ref{fig_cycle}). (i) Isentropic compression $A\rightarrow B$: the polariton is initiated at the temperature $T_1\sim T_a$ and frequency $\Omega_1$ (mode $k_1$), which is sweeped until the value $\Omega_2$ (mode $k_2$) with the help of the piezoelectric cells. The volume of a mode can be defined as $V= {\cal S}/ k$, such that the relation $k_2/k_1$ can be expressed in terms of a volume ratio as $V_1/V_2$, as in Fig.~\ref{fig_cycle}. This stroke must be fast enough such that the polariton number $N_k^{(L)}=\langle A_k^\dagger A_k\rangle=\vert u_k \vert^2 n_k^{(a)}+\vert v_k \vert^2 n_k^{(b)}$ is kept constant at its initial value $N_1\equiv N_{k_1}^{(L)}\sim n_{a}$, but slow enough such that jumps to the upper polariton mode are suppressed. Therefore, the duration $\tau_1$ of the stroke must satisfy the constraint $\Lambda\gg 1/\tau_1\gg \Gamma$, such that strong coupling holds. (ii) Isochoric heating $B\rightarrow C$: the volume of the system is kept constant at the value $V_2$ and is allowed to thermalize with the hot source at $T_2\sim T_b$. This process has a duration $\tau_2\sim 1/\Gamma \gg \tau_1$, during which the polariton number increases from $N_1$ to its final value $N_2\equiv N_{k_2}^{(L)}\sim n_{b}$ (see Fig.~\ref{fig_dispersion} b)). (iii) Isentropic expansion $C\rightarrow D$: this stroke consists in the reversed sweep of the frequency (volume) from the value $\Omega_2$ ($V_2$) to its initial value $\Omega_1$ ($V_1$). The duration of this process is $\tau_3 \sim \tau_1$. (iv) Isochoric cooling $D\rightarrow A$: in this transformation, with duration $\tau_4\sim \tau_2$, the system expels heat by thermalizing with the cold source, at the constant volume $V_1$. 

In the conditions above, the energy of each point of the cycle can be computed as $\langle H \rangle_A=\hbar \Omega_1 N_1$, $\langle H \rangle_B=\hbar \Omega_2 \mathcal{Q}^*_1 N_1$, $\langle H \rangle_C=\hbar \Omega_2 N_2$ and $\langle H \rangle_D=\hbar \Omega_1 \mathcal{Q}^*_2 N_2$, where $\mathcal{Q}^*_{1,2}$ are parameters measuring the adiabaticity of the isentropic strokes (i) and (iii). Adiabatic (non-adiabatic) transformations satisfy $\mathcal{Q}^*_j=1$ ($\mathcal{Q}^*_j>1$) \cite{husimi, lutz_pre}. The efficiency of the machine can thus be defined as $\eta=W/Q_h$, where $W=\langle H \rangle_C- \langle H \rangle_D+\langle H \rangle_A-\langle H \rangle_B$ is the total work output and $Q_h=\langle H \rangle_C- \langle H \rangle_B$ is the heat received from the hot source, yielding  
\begin{equation}
\eta=1-\frac{Q_c}{Q_h}=1-\frac{\Omega_1}{\Omega_2}\frac{N_2 \mathcal{Q}^*_2-N_1}{N_2-N_1 \mathcal{Q}^*_1}.
\label{eq_eff}
\end{equation}
Eq.~(\ref{eq_eff}) generalizes the result of Ref. \cite{zhang}, obtained in an optomechanical setup. Since heat is absorved from the reservoir, $Q_h>0$, and flows into the cold reservoir, $Q_c<0$, the following conditions must be satisfied, $\mathcal{Q}_1^*\leq N_2/N_1$, $\mathcal{Q}_2^*\geq N_1/N_2$. This condition is achieved for experimentally accessible parameters ($N_1/N_2 \sim 0.127$, see Fig.~\ref{fig_dispersion}). Without loss of generality, we now assume that both compression and expansion have the same duration, $\tau_1=\tau_3\equiv  \tau$, which implies $\mathcal{Q}_1^*=\mathcal{Q}_2^*\equiv \mathcal{Q}^*$. Thus, the upper bound for the performance is achieved for the condition of work reversion, i.e. $W<0$, which corresponds to the Carnot efficiency $\eta_{\rm Carnot}=1-T_1/T_2$. Moreover, in the limit $k_B T_2\gg \hbar \Omega_2$, we can {\it upper} bound the maximum efficiency as $\eta_{\max} \leq \eta_+ \leq \eta_{\rm Carnot}$, where 
\begin{equation}
\eta_+ =1-\frac{T_1}{T_2} \left[2 \mathcal{Q}^*\left(\mathcal{Q}^*+\sqrt{{\mathcal{Q}^*}^2-1}\right)-1\right].
\label{Eq_Eff}
\end{equation}
An important feature of QHEs is the output work and the efficiency at maximum power, $\eta_*$. Indeed, in finite-time thermodynamics (FTT), there is a tradeoff between maximum power and maximum efficiency, at which the output power vanishes. For a quantitative analysis, we maximize the total work $W=\hbar \Omega_2(N_2-\mathcal{Q}^* N_1)+\hbar \Omega_1(N_1-\mathcal{Q}^* N_2 )$ for a quasi-adiabatic cycle consisting of a frequency modulation of $\Omega (t)=\Omega_1+(\Omega_2-\Omega_1) t/\tau$. Correspondingly, the adiabaticity parameter can be determined according to Ref.~\cite{husimi, lutz_pre} and reads $\mathcal{Q}^*\simeq 1 + \alpha$, where $\alpha=(\Omega_2-\Omega_1)^2/(8\tau^2\Omega_2^4)\ll 1$. Maximization with respect to $\Omega_1/\Omega_2$ allows us to {\it lower} bound the efficiency at maximum power by
\begin{equation}
\eta_-=\frac{1+\theta-2\mathcal{Q}^*\sqrt{\theta}}{1-\mathcal{Q}^*\sqrt{\theta}}, \quad \theta=T_1/T_2.
\end{equation} 
If the isentropic strokes are performed adiabatically, $\alpha \rightarrow 0$, the latter reduces to the Curzon-Ahlborn (CA) efficiency $\eta_{\rm CA}=1-\sqrt{T_1/T_2}$. The efficiency in Eq.~(\ref{eq_eff}) and the total work are respectively depicted in Fig.~\ref{fig_Eff} a) and b) for adiabatic (${\cal Q}^*=1$) and the quasi-adiabatic (${\cal Q}^*=1.005$) strokes, corresponding to a sweeping time $\tau =\infty$ and $\tau\simeq 2.5/\Omega_1\sim 0.1 \mu$s, compatible with the response time of piezoelectric cells.

\par{\it Quantum Refrigerator.}
%
\begin{figure}[t!]
\includegraphics[scale=0.33]{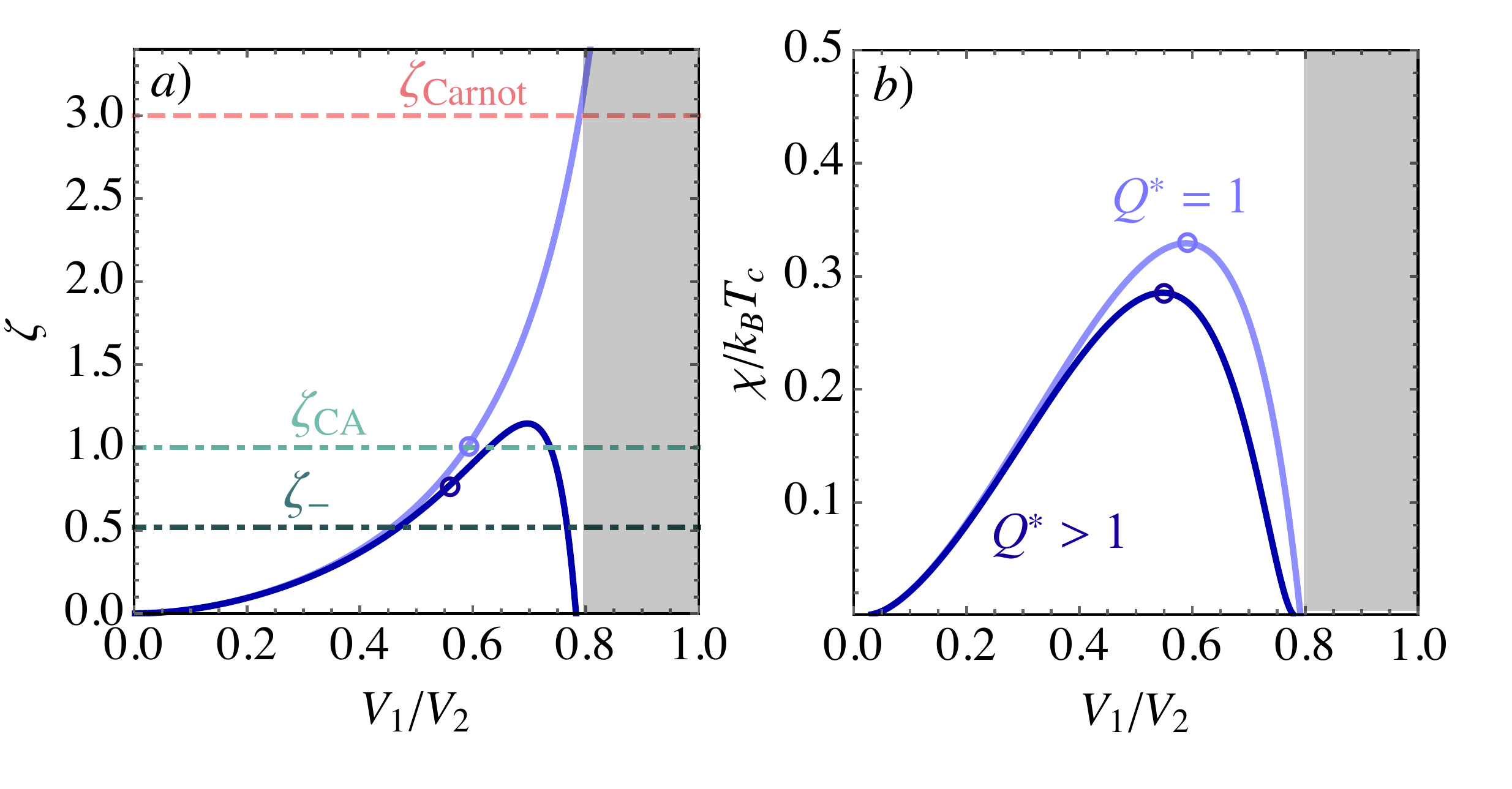}
\caption{(color online) a) Coefficient of performance (CoP) of a quantum refrigerator, and b) Figure-of-merit of the refrigerator, for the same parameters of Fig. \ref{fig_Eff}. The performance of the machine is upper-bounded by the Carnot CoP. The circles represent the compression ratio for which the power is maximum and the shadowed area represents the region of heat reversion ($Q_c<0$). The CoP at maximum power is lower-bounded by the Curzon-Ahlborn CoP for the adiabatic case and by the value $\zeta_-$ discussed in the text for quasi-adiabatic case. $T_a=7.5$ mK and $T_b=10$ mK.}
\label{fig_CoP}
\end{figure}
The inversion of the cycle in Fig.~\ref{fig_cycle} allows the lower polariton mode to fuel a quantum refrigerator. The coefficient of performance is then determined as the ratio of the heat extracted from the cold source to the work consumed, $\zeta = Q_c/W=Q_c/(Q_h-Q_c)$. Repeating the analysis performed for the QHE, we obtain $Q_c=\hbar \Omega_1 \left(N_1- \mathcal{Q}^* N_2\right)$ and the heat delivered to the hot source is $Q_h=\hbar \Omega_2 \left({\cal Q}^*N_1-N_2\right)$, which yields 
\begin{equation}
\zeta=\frac{\Omega_1 \left( N_1- \mathcal{Q}^*N_2 \right)} {\Omega_2 \left(\mathcal{Q}^* N_1-N_2 \right)-\Omega_1 \left(N_1 - \mathcal{Q}^*N_2 \right)}.
\end{equation}
The FTT analysis of the performance of a quantum refrigerator is more involved, as the maximization of the cooling power (or, equivalently, the minimization of the work input) does not result in a temperature-dependent bound for $\zeta$ \cite{yan, velasco, abah}. In fact, by maximizing the heat-pumping power one simultaneously minimizes $\zeta$ to zero. Alternatively, a suitable figure-of-merit is defined as the product of the extracted heat and the coefficient of performance, $\chi=Q_c \zeta $ \cite{allahverdyan} (see Fig.~\ref{fig_CoP} b)). A FTT {\sl lower} bound $\zeta_{-}$ for performance of the refrigerator can thus be obtained in the limit $\hbar \Omega_2 \ll k_B T_2$, for which we obtain $\chi \simeq x^2\left(\theta -{\cal Q}^*\right)/\left(\frac{\theta {\cal Q}^*-x}{\theta-{\cal Q}^*}-x\right)$, with $x=\Omega_1/\Omega_2$. Optimization with respect to $x$ yields
\begin{equation}
\zeta_- = \zeta _{\rm CA}- \frac{2 \theta+\sqrt{1-\theta}}{\left( 1- \theta \right)^2}\left({\cal Q}^*-1\right),
\label{CoP_optimal}
\end{equation}
where $\zeta_{\rm CA}= 1/\sqrt{1-\theta}-1$ is the classical CA coefficient of performance \cite{yan2}. Semi-classical estimates of Eq.~(\ref{CoP_optimal}) can be worked out in the case where the cold source is in the quantum regime provided the replacement $\theta \rightarrow \theta_{\rm sc}=\hbar \Omega_1 \coth (\hbar\Omega_1/2k_BT_1)/(2k_B T_2)$, and in the full quantum regime with $\theta\rightarrow \theta_{\rm q}=N_1/N_2=\coth(\hbar \Omega_1/2 k_B T_1)/\coth(\hbar \Omega_2/2 k_B T_2)$ \cite{abah}. These results are illustrated in Fig.~\ref{fig_CoP}, where we can observe that the coefficient of performance at maximum power is bounded as $\zeta_- \leq \zeta_*\leq \zeta_{\rm Carnot}$, where $\zeta_{\rm Carnot}=T_1/(T_2-T_1)$. 

{\par \it Conclusion.}
We have proposed a quantum thermal machine - working either as a quantum heat engine or as a quantum refrigerator - based on the vacuum forces between a graphene and a gold nanoresonator. The machine working fluid is the lower polariton resulting from the hybridization of the graphene- and gold-like flexural modes, interacting via Casimir (vacuum) forces. With the help of piezoelectric cells, we select and sweep the hybridized mode frequency, alternating the thermal contact with the two reservoirs. The operation of our machine depends on the strong-coupling condition, a feature that is a hallmark of nanomechanical systems at cryogenic temperatures, as the decaying rate is much weaker than the avoided-crossing frequency. Due to a Rabi frequency of a few MHz, it is possible to perform quasi-adiabatic strokes of duration $\sim 0.1 \mu$s, thus suppressing transitions to the upper polariton branch. We further observe that high performance is compatible with a finite-time thermodynamical analysis within a experimentally accessible set of parameters. In particular, we obtain typical efficiencies at maximum power of $\sim 71\%$, for the system operating as a heat engine, and a coefficient of performance of $\sim 0.76$ for the refrigeration cycle. Since the performance of our cycle is proportional to the ratio between the final and initial frequencies, our proposal is specially advantageous. This happens because our resonators have very different bending stiffnesses. These features suggest that our thermodynamical cycle may be relevant for temperature control (or temperature measurements) in experiments of suspended films. Particularly important are the technologies involving graphene at cryogenic temperatures \cite{song2014}, for which both the electric and thermal resistivity mostly depends on the flexural phonons \cite{jiang2015}. 
\par
The authors thank the support from Funda\c{c}\~{a}o para a Ci\^{e}ncia e a Tecnologia (Portugal), namely through programmes PTDC/POPH/POCH and projects UID/EEA/50008/2013, IT/QuSim, ProQuNet, partially funded by EU FEDER, and from the EU FP7 project PAPETS (GA 323901). Furthermore, MP acknowledges the support from the DP-PMI and FCT (Portugal) through scholarship SFRH/BD/52240/2013, and HT thanks acknowledges the support from FCT (Portugal) through scholarship SFRH/BPD/110059/2015.  

\section{Supplemental Material}
\subsection{Casimir interaction at cryogenic temperatures}

The Casimir force between two electrically neutral objects with or without permanent electric or magnetic moments macroscopic bodies originates from the zero-point fluctuations of the electromagnetic field \cite{casimir, lifshitz}. This interaction can be derived in many different ways, here we will follow the formalism  where the interaction is described in terms of the electromagnetic normal modes, transverse electric (TE) or $s$-polarized and transverse magnetic (TM) or $p$-polarized of the system \cite{PRB.85.195427.2012}. For the case of zero temperature, the Casimir energy per unit area between two planar surfaces is given by
\begin{eqnarray}
E (d) = \frac{\hbar}{4 \pi^2} \sum_{j=\mathrm{TE,TM}}\int_{0}^\infty k dk \int_0^\infty d \xi \ln \pr{f_k^j \pc{i \xi}},
\label{eq:EnergyT0}
\end{eqnarray}
with $k$ being the projection of the wave vector on the plane of the surface and $\xi=i\omega$ where $\omega$ is the frequency of the TE and TM modes. For geometries consisting of three regions (that we take to be vacuum) and two interfaces $1 \vert 2 \vert 3$, as in Fig. 1 of the manuscript, the mode condition function is written as
\begin{eqnarray}
f_k^j = 1 - \exp \pc{-2 \kappa_2 k d} r_{21}^j r_{23}^j,
\end{eqnarray}
where $r_{mn}$ is the reflection coefficient for a wave on the interface between medium $m$ and $n$ from the $m$ side, $d$ is the thickness of region 2, and $ \kappa_n = \sqrt{1+\varepsilon_n \pc{i \xi} \pc{\xi/c k}^2}$, where $\varepsilon_n (\omega)$ is the dielectric function of the medium $n$ and $c$ is the speed of light. 

As Eq.~(\ref{eq:EnergyT0}) is valid for zero temperature only, to treat more realistic case of media at finite temperature one needs to consider fluctuating currents inside the objects, including both zero point and thermal fluctuations. At finite temperature the second integral in Eq.~(\ref{eq:EnergyT0}) is replaced by a summation over the discrete Matsubara frequencies 
\begin{eqnarray}
\xi_l = \frac{2 \pi k_B T l}{\hbar}, \quad l=0,1,2,\dots,
\end{eqnarray}
such that
\begin{eqnarray}
E (d, T)  = \frac{k_B T}{2 \pi} \sum_{j=\mathrm{TE,TM}} \sum^{\infty}_{l=0}{}^{'} \int_{0}^\infty k dk  \ln \pr{f_k^j \pc{i \xi_l}}.
\label{eq:EnergyTfinite}
\end{eqnarray}
The prime on the summation sign indicates that the term with $l=0$ is reduced by a factor of two \cite{sofiaPRA2013}. 

Both Eqs.~(\ref{eq:EnergyT0}) and (\ref{eq:EnergyTfinite}) need to account both the transverse magnetic (TM) and the transverse electric (TM) modes. The TM reflection coefficient of graphene can be expressed via the polarization tensor $\Pi_{00}$ as \cite{PRB89.125407.2014}
\begin{eqnarray}
r_{\text{TM}}\pc{i \xi_l, k} = \frac{\kappa_2 \Pi_{00} \pc{i \xi_l, k}}{2 \hbar k + \kappa_2 \Pi_{00}\pc{i \xi_l, k} },
\label{eq:rTM}
\end{eqnarray} 
which, for undopped graphene, reads
\begin{align}
& \Pi_{00}\pc{i \xi_l, k} = \frac{\pi \hbar \alpha k^2}{f \pc{\xi_l, k}} + \frac{8 \hbar \alpha c^2}{v²_F} \int_0^1 dx \left\lbrace \frac{k_B T}{\hbar c} \right. \nonumber \\
& \ln \pr{1 + 2 \cos \pc{2 \pi l x} e^{- \theta_T \pc{\xi_l, k, x}} +e^{- 2 \theta_T \pc{\xi_l, k, x}}}  \nonumber \\
&-\frac{\xi_l}{2 c} (1-2 x) \frac{\sin (2 \pi l x)}{\cosh \pr{\theta_T \pc{\xi_l, k, x}} + \cos (2 \pi l x) } \nonumber \\
&\left. +\frac{\xi^2_l \sqrt{x(1-x)}}{c^2 f\pc{\xi_l,k}} \frac{\cos (2 \pi l x) + e^{- \theta_T \pc{\xi_l, k, x}}}{\cosh \pc{ \theta_T \pc{\xi_l,k, x}} +\cos (2 \pi l x)} \right\rbrace,
\label{eq_polarization_tensor}
\end{align}
with $\alpha = e^2 / (\hbar c)$ standing for the fine-structure constant and $v_F = 8.73723 \times 10^5$~m/s for the Fermi velocity where it was defined
\begin{eqnarray}
f \pc{\xi_l,k} &=& \sqrt{\frac{v^2_F}{c^2} k^2 + \frac{\xi^2_l}{c^2}}, \\
\theta_T \pc{\xi_l, k, x} &=& \frac{\hbar c}{k_B T} f \pc{\xi_l, k} \sqrt{x(1-x)}.
\end{eqnarray}
The TE reflection coefficient, on the other hand, takes the form \cite{PRB89.125407.2014}
\begin{eqnarray}
r_{\text{TE}}\pc{i \xi_l, k} = - \frac{\Pi_{\text{tr}} \pc{i \xi_l, k} - \kappa_2^2 \Pi_{00} \pc{i \xi_l, k}}{2 \hbar \kappa_2 k + \Pi_{\text{tr}} \pc{i \xi_l, k} -  \kappa_2^2 \Pi_{00} \pc{i \xi_l, k} },
\label{eq:rTE}
\end{eqnarray} 
where $\Pi_{\text{tr}}$ is given by 
\begin{align}
&\Pi_{\text{tr}} \pc{i \xi_l, k} = \Pi_{00} \pc{i \xi_l, k} + \frac{\pi \hbar \alpha}{f \pc{i \xi_l, k} } \pr{f^2 \pc{i \xi_l, k} +\frac{\xi^2_l}{c^2}} \nonumber \\
&+ 8 \hbar \alpha \int_0^1 dx \left \lbrace \frac{\xi_l}{c}  \right. \frac{ (1-2x) \sin (2 \pi l x)}{\cosh \pr{\theta_T \pc{i \xi_l, k,x}} + \cos (2 \pi l x)} \nonumber \\
&-\frac{\sqrt{x(1-x)}}{f\pc{i \xi_l, k}} \pr{f^2\pc{i \xi_l, k} + \frac{\xi^2_l}{c^2}} \nonumber \\
&\left. \times \frac{\cos (2 \pi l x) + e^{- \theta_T \pc{i \xi_l, k,x}}}{\cosh \pr{\theta_T \pc{i \xi_l, k,x}} + \cos (2 \pi l x)} \right\rbrace .
\label{eq_zero_pol}
\end{align}
At zero temperature, the discrete frequencies $\xi_l$ are replaced by the continuous $\xi$ and $\theta_T \pc{\xi_l,k,x} \to \infty$, such that
\begin{eqnarray}
\Pi_{00} \pc{i \xi, k} = \frac{\pi \hbar \alpha k^2}{f (\xi, k)}.
\end{eqnarray} 
Then, the reflection coefficients of graphene simply read \cite{PRB.85.195427.2012}
\begin{eqnarray}
r_{\text{TM}} \pc{i \xi, k} = \frac{\pi e^2 \sqrt{c^2 k^2 + \xi^2}}{2 \hbar c \sqrt{v_F^2 k^2 + \xi^2} + \pi e^2 \sqrt{c^2 k^2 + \xi^2}}
\label{eq:rTMT0}
\end{eqnarray}
and
\begin{eqnarray}
r_{\text{TE}} \pc{i \xi, k} = - \frac{\pi e^2 \sqrt{v^2_F k^2 + \xi^2}}{2 \hbar c \sqrt{c^2 k^2 + \xi^2} + \pi e^2 \sqrt{v_F^2 k^2 + \xi^2}}.
\end{eqnarray}
\begin{figure}
\includegraphics[scale=0.59]{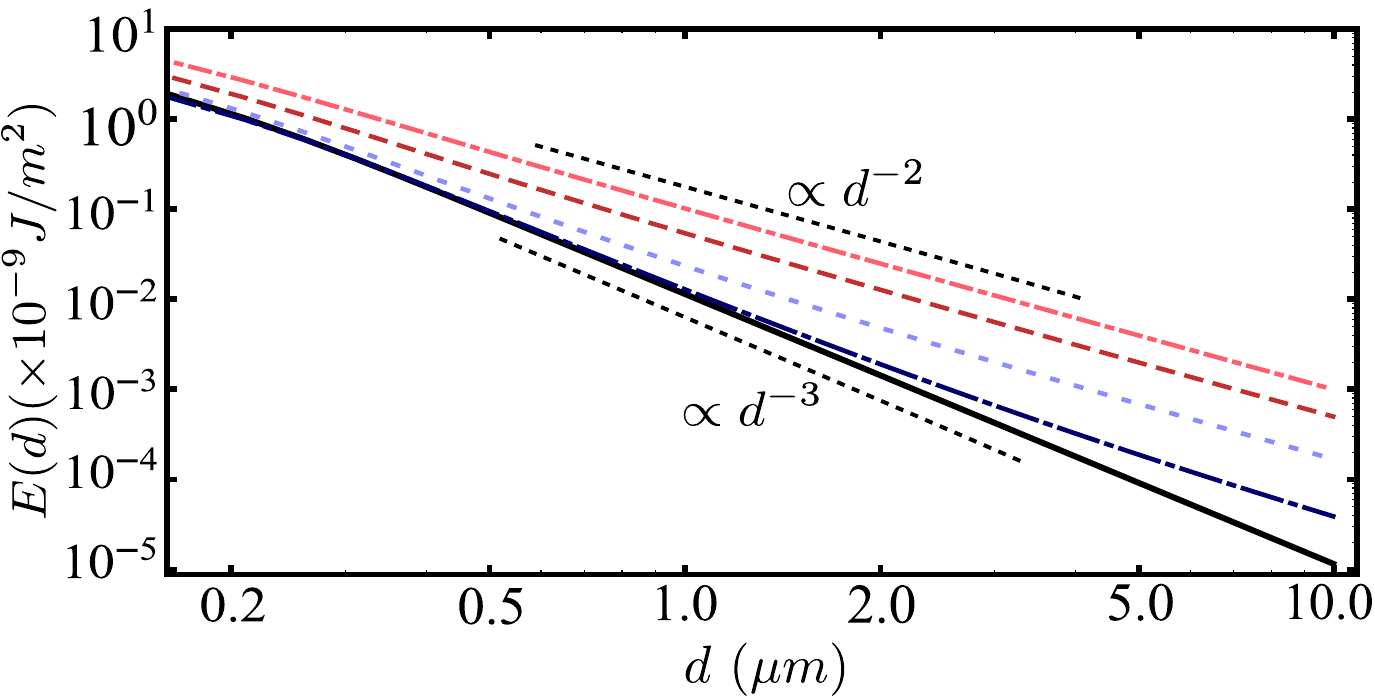}
\caption{(color online) Casimir energy for different values of the temperature. From top to bottom, $T=\{300, 150, 50, 10\} $ K. The solid line is the quantum ($T=0$) case. \label{Fig_forces}} 
 \end{figure}
As stated in the main text, we are interested in computing the Casimir interaction between thin gold films and graphene. In Ref.~\cite{EurPhysJ.B86:43.2013}, it is shown that the Casimir forces for ultrathin films is $\sim 20\%$ larger then the corresponding one for bulk gold. Also, when the film thickness increases, the use of isotropic bulk dielectric function becomes an increasingly good approximation. As such, we considered the dielectric function of gold to be a good approximation in our case. The reflection coefficients for 2D gold sheet can be calculated by matching the dyadic Green function of free space and its derivatives on either side of a two-dimensional conducting sheet. For those cases the reflections coefficients are reduced to \cite{PRB.85.195427.2012}
\begin{eqnarray}
r_{\text{TM}}^\text{gold} &=& \frac{\kappa_{2} \alpha_{\parallel} (k, i \xi)}{1+\kappa_{2} \alpha_{\parallel} (k_, i \xi)}, \\
r_{\text{TE}}^\text{gold} &=& \frac{(i \xi / c k)^{2}  \alpha_{\perp} (k, i\xi) }{\kappa_{2} -(i \xi / c k)^{2}  \alpha_{\perp} (k, i \xi)}.
 \label{eq:rgold}
\end{eqnarray}
The dielectric function of a 2D system is given by $\varepsilon (k, \omega) = 1 + \alpha (k, \omega)$, where $\varepsilon \pc{\omega} = 1 - \frac{\omega_p^2}{i \nu \omega + \omega^2}$, with $\omega_p = 1.37\times 10^{16}$~rad/s and $\nu=4.12\times10^{13}$~rad/s. A numerical fit to Eq.~(\ref{eq:EnergyT0}) $E(d)\sim C_3/d^3$, with $C_3\simeq -1.14 \times 10^{-11}$ Jm$^2$ (solid line in Fig.~\ref{Fig_forces}). 
\subsection{Casimir interaction with graphene systems at finite temperature}

The effect of the temperature on the dielectric function of graphene is negligible, for all Matsubara frequencies but for $l = 0$. Therefore, we safely make use of the zero-temperature result Eq.~(\ref{eq_zero_pol}) for all $l$ except $l=0$. 

The Casimir energy per unit area for the gold-graphene interface of the present work is depicted in Fig.~\ref{Fig_forces} for different temperatures. At $T=300$ K ($T=70$ K), we obtain the power-law behaviour $E(d)\simeq C_2/d^{2}$, with $C_2=-9.91 \times 10^{-11}$ ($C_2=-2.40 \times 10^{-11}$) Jm. By decreasing the temperature, we observe a crossover from $\sim d^{-2}$ to the $\sim d^{-3}$ power laws. Ultimately, at the cryogenic temperatures of few mK considered in this work, deviations from the quantum result are negligible. This fact rules out the need of including out-of-equilibrium corrections, which are only relevant at near-to-room temperatures. In fact, for the general case $T_a \neq T_b \neq 0$, it is possible to use the additivity property of the thermal energy which can be written, in general, as the sum of two contributions \cite{PRA77.022901.2008,PRL95.113202.2005}, such that the Casimir energy $E^\text{neq}_\text{th} \pc{T_a, T_b, d} $ for the two sheets at different temperatures is given by
\begin{align}
E^\text{neq}_\text{th} \pc{T_a, T_b, d} &= \frac{E \pc{T_a, d} + E \pc{T_b, d}}{2}. \nonumber\\
 &+ E^\text{neq} \pc{T_a, T_b, d} \label{eq:OutOfEquilibrium}
\end{align}
and the out-of-equilibrium correction reads \cite{PRA92.032116.2015}
\begin{align}
&E^\text{neq}_\text{th} \pc{T_a, T_b, d} = \frac{\hbar}{4 \pi^2} \int_0^\infty d \omega \pr{n \pc{\omega,T_a}-n\pc{\omega,T_b}}\nonumber \\
& \int_0^\infty k dk \sum_{j=\mathrm{TE,TM}} \text{Im} \pr{\ln \pc{1-e^{2 i d \sqrt{\omega^2 /c^2 - k^2}} r_{21}^j r_{23}^j}} \nonumber \\
&\times \left\lbrace \theta \pc{\frac{\omega}{c} - k} \frac{\md{r_{23}^j}^2-\md{r_{21}^j}^2}{1-\md{r_{21}^j r_{23}^j}^2} \right. 
\left. +
\theta \pc{k - \frac{\omega}{c}}  \frac{\text{Im}\pr{r_{21}^j r_{23}^{j*}} }{\text{Im} \pr{r_{21}^j r_{23}^j} } \right\rbrace,
\label{eq_out_eq}
\end{align}
with $\theta (x)$ being the unit step function and and $n \pc{\omega, T} = \pr{\exp \pc{\hbar \omega / (k_B T)} -1}^{-1}$ the Bose-Einstein distribution. Eq.~(\ref{eq_out_eq}) vanishes identically if the sheets have identical reflection coefficients. 

\subsection{Kirchhoff-Love Theory and the Polariton Dispersion}

The mechanical vibrations of the system are described by the Kirchhoff-Love plate theory \cite{amorimPRB}. The Lagrangian density of the $j-$plate ($j=a, b$) can be written as $\mathcal{L}_j=\rho_j \dot {\bm \eta}_j^2/2-\mathcal{H}_j$, where $\rho_j$ is the areal mass density and ${\bm \eta}_j({\bm x})=\eta_j ({\bm x})\mathbf{e}_z$ is a continuous vector field describing the vertical (out-of-plane) displacement of the $j$th membrane at position ${\bm x}=(x,y)$. The energy density can thus be expressed as
\begin{equation} 
\mathcal{H}_j=\frac{1}{2}D_j\left(\nabla^2{\bm \eta}_j\right)^2+ \gamma_{x,j} \left(\frac{\partial {\bm \eta}_j}{\partial x}\right)^2+\gamma_{y,j} \left(\frac{\partial {\bm \eta}_j}{\partial y}\right)^2.
\label{Eq_hamilton1}
\end{equation}
Here, $D_j$ is the bending stiffness, $\gamma_{x,j}$ ($\gamma_{y,j}$) is clamping the tension along the $x$ ($y$) direction to the clamping with the substrate and $\nabla^2{\bm \eta}_j$ is the local curvature. From the Euler-Lagrange equations for displacements of the form ${\bm \eta}_j({\bm x})\sim e^{i {\bm k}\cdot {\bm x}-\omega_k^{(j)}t}$, we obtain bare-mode frequencies $j=(a, b)$ as
\begin{equation}
\omega_k^{(j)}=\sqrt{D_j k^4+\gamma_{x,j} k_x^2+\gamma_{y,j} k_y^2}, \label{eq_modes}
\end{equation} 
For simplicity, we consider the case of isotropic clamping, such that $\gamma_{x,j}=\gamma_{y,j}=\gamma_j$. For gold membranes, the stiffness is a function of the thickness $\delta$ as $$ D_a=\frac{1}{12}\frac{E_a}{1-\nu_a^2}\delta^3,$$ where $E_a=79$ GPa is the (bulk) Young's modulus and $\nu_a=0.4$ is the Poisson strain coefficient. Here, we have chosen $\delta=10$ nm, such that the bulk reflection coefficients used in the computation of the vacuum forces are valid. For graphene, $D_b\sim 1.5$ eV is experimentally accessible \cite{lindahl2012}. Also, we have considered plates of size $L=15~\mu$m, such that $d\ll L$, enabling them to be considered as infinite and therefore it is safe to perform the integral over the wavevector $k$ in Eq.~\eqref{eq:EnergyT0}. The frequency $\omega_0^{(j)}$ of the fundamental mode can be obtained by evaluating Eq.~\eqref{eq_modes} at $k=2\pi/L\equiv k_0$. The corresponding zero-point displacement is thus obtained from the relation $$ h_0^{(j)}=\sqrt\frac{\hbar}{M_j\omega_0^{(j)}},$$ with $M_j=\mathcal{S} \rho_j$ standing for the resonator mass and $\mathcal{S}=L^2$. A summary of the experimentally accessible parameters can be found in Table~\ref{Table_parameters}. The single-particle Hamiltonian can then quantized via the replacement
\begin{equation}
{\bm \eta}_j({\bm x})=\frac{1}{\mathcal{S}} \sum_{{\bm k},\sigma} h_{{\bm k}}^{(j)}e^{i {\bm k } \cdot {\bm x}}\left(c_{\bm k}^{(j)}+c_{\bm k}^{(j) \dagger} \right)e_{\sigma},
\end{equation}
which yields $H_0=\hbar \sum_{k,\sigma}\left( \omega_k^{(a)}a_{k, \sigma}^\dagger a_{k,\sigma}+\omega_k^{(b)}b_{k, \sigma}^\dagger b_{k,\sigma}\right)$, as discussed in the main text. 

In the following, we compute the correction to the Casimir potential Eq.~(\ref{eq:EnergyTfinite}) due to the flexural modes. Making use of the Proximity Force Approximation \cite{thorsten2001}, we obtain a local correction at position ${\bm x}$ due to a ripple in the $j$th membrane as
\begin{eqnarray}
\mathcal{U}_{jj}({\bm x}; d)&\equiv&  ~E(d-\eta_j({\bm x}))\nonumber\\
&\simeq&  E(d)-\eta_j({\bm x})\left. \frac{ d E(z)}{d z} \right\vert_{z=d},
\end{eqnarray}
valid the limit of small displacements $\eta_j(x)\ll d$. Following Ref. \cite{thorsten2001}, we obtain the explicit correction in the short wavelength limit $k d\gg 1$
\begin{equation}
\mathcal{U}_{jj}({\bm x}; d)\simeq \frac{C_3}{d^3}\left[ 1+k\frac{\eta_j^* \eta_j}{ d}\right]. 
\label{eq_correction}
\end{equation}
\begin{figure}
\includegraphics[scale=0.9]{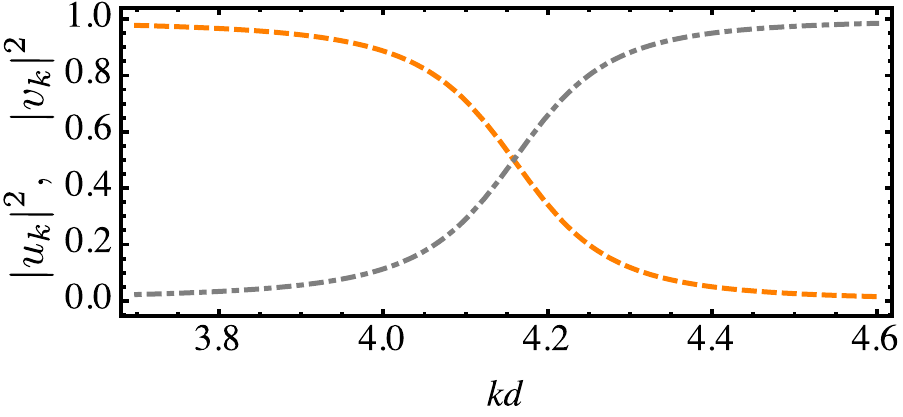}
\caption{(color online) Hopfield coefficients measuring the fraction of gold (orange, dashed line) and graphene (gray, dotdashed line) in the lower polariton mode $A_k$.} 
\label{Fig_Hopfield}
 \end{figure}
This is at the origin of the shift of the bare frequencies in Eq.~(\ref{eq_modes}), as discussed in the main text.  A reasonable approximation to the potential correction due to ripples in {\it both} plates is taken into account by replacing Eq.~(\ref{eq_correction}) by $\mathcal{U}_{ij}\simeq \frac{C_3}{d^3}\left[ 1+k\frac{\eta_j^* \eta_i}{ d}\right]$ and its complex conjugate. The latter is responsible for the coupling between the modes. The quantization procedure introduced previously and discussed in the text finally leads to Eq. (4) of the manuscript, which can then be diagonalized with the help of a Hopfield-Bogoliubov transformation, by defining the operators $A_k=u_k a_k+v_k b_k$ and $B_k=v_k b_k-u_k a_k$. The Hopfield coefficients, representing the fraction of gold and graphene, are given by in terms of the polariton frequencies $\Omega_k^{(L)}$ and $\Omega_k^{(U)}$  as \cite{guillaume_book}
\begin{equation}
\begin{array}{c}
\vert u_k\vert^2=\frac{\omega_k^{(a)} \Omega_k^{(U)}-\omega_k^{(b)}\Omega_k^{(L)}}{\left(\omega_k^{(a)} +\omega_k^{(b)} \right) \sqrt{4 g_k^2+\left(\omega_k^{(a)} +\omega_k^{(b)} \right)^2}},\\\\
\vert v_k\vert ^2=\frac{\omega_k^{(b)} \Omega_k^{(U)}-\omega_k^{(a)}\Omega_k^{(L)}}{\left(\omega_k^{(a)} +\omega_k^{(b)} \right) \sqrt{4 g_k^2+\left(\omega_k^{(a)} +\omega_k^{(b)} \right)^2}}.
\end{array}
\end{equation}
The thermodynamical cycle will operate in the lower polariton mode. The corresponding number of excitations at a given mode $k$, $N_k^{(L)}=\langle A_k^\dagger A_k \rangle$ can then be determined as
\begin{equation}
N_k^{(L)}=\vert u_k\vert^2 n_k^{(a)}+\vert v_k\vert^2 n_k^{(b)},
\end{equation}
where $n_k^{(j)}=\left(\exp({\hbar \beta_j \omega_k^{(j)}})-1\right)^{-1}$ is the Bose-Einstein distribution, $\beta_j=1/k_B T_j$ and $T_j$ is the temperature of the membrane $j$. At cryogenic temperatures($T_a=1$ mK and $T_b=10$ mK), $N_k^{(L)}\lesssim 1$ for some modes $k$, suggesting that our polariton-based thermomachine is working in the quantum regime. 

\section{Polariton Coherence and Lifetime}

Although the polariton mode results from the coherent superposition between the flexural modes $a$ and $b$, incoherence processes are also present. The polariton decay rate $\Gamma$ encompasses two main processes, namely the flexural phonon lifetime $\Gamma_{\rm ph}$ and the polariton-phonon decay $\Gamma_{{\rm pol} \rightarrow {\rm ph}}$, 
\begin{equation}
\Gamma=\Gamma_{\rm ph}+ \Gamma_{{\rm pol} \rightarrow {\rm ph}}.
\end{equation}
At cryogenic temperatures, the phonon-phonon scattering is very weak, which means that the flexural modes are long-lived at the relevant time scales, $\Gamma_{\rm ph}\sim 0$ \cite{jiang2015}. The polariton-phonon decay rate, in its turn, can be estimated with the help of Fermi's golden rule 
$$\Gamma_{{\rm pol} \rightarrow {\rm ph}}=\frac{2\pi}{\hbar} \sum_{k,q} \vert \langle k \vert H_{\rm int} \vert q \rangle \vert^2 \delta\left(\omega_k^{(a)}-\omega_q^{(a)}\right),$$ where $H_{\rm int}=\sum_k g_k a_k^\dagger b_k+ {\rm h. c.}$ is the interaction Hamiltonian. The initial and the final states are chosen in such a way that only inelastic processes are taken into account, and are thus respectively given by $$\vert k \rangle = \vert n_{k}^{(a)}, n_{k}^{(b)}\rangle, \quad {\rm and}$$ $$\vert q \rangle=\frac{1}{\sqrt{2}}\left( \vert n_{q}^{(a)}+1,n_{q}^{(b)}-1\rangle+ \vert n_{q}^{(a)}-1,n_{q}^{(b)}+1 \rangle\right).$$ Some simple algebra yields
\begin{align*}
\Gamma_{{\rm pol} \rightarrow {\rm ph}}&= \frac{\pi}{\hbar}\sum_k \vert g_k\vert^2 (1+n_{k}^{(a)})(1+ n_{k}^{(b)}) \delta(\omega_k^{(a)}-\omega_k^{(b)}),\\
&\simeq \frac{\pi \Lambda^2}{\omega_c^2},
\end{align*}
where we have used $g_k=g_k^*$. For the parameters of the quantum heat engine (QHE) and the quantum refrigerator (QR) (see Table~\ref{Table_parameters}), we estimate $\Gamma$ to range from $0.08 \Lambda$ to $0.2 \Lambda$. 

\begin{widetext}
\begin{table*}[t]
\label{Table_parameters}
\caption{Parameters used in the design of a quantum thermal machine based on the flexural modes of a graphene-gold nanoresonator interface}
\begin{center}
 \begin{tabular}{c c c c c c c c c c} 
 \hline
 \hline
              &  $\rho$ (mg/m$^{-2}$)    & $\delta$ (nm)  & $L$ ($\mu$m)& $\gamma$ (nN/$\mu$m) & $D$ (eV) & $h_0$ (fm) & $\omega_0$ (MHz) & T (mK) [QHE] & T (mK) [QR]\\ 
 \hline
Graphene  & $ 0.761$   &  0.3                  &    15                                          &          0.22                        &     1.5         &  5.3          &       7.2        & 10 & 10              \\
Gold           & $193$      &    10                 &    15                                          &          0.10                         &     48 983   &  1.1          &      2.6       & 1.0 & 7.5		\\
 \hline
 \hline
\end{tabular}
\end{center}
\end{table*}
\end{widetext}

\end{document}